\documentclass[aps,showpacs,twocolumn,superscriptaddress]{revtex4}

\usepackage{amsmath}
\usepackage{amssymb}
\usepackage{amscd}
\usepackage{wasysym}
\usepackage[ansinew]{inputenc}
\usepackage[T1]{fontenc}
\usepackage{ae,aecompl}

\usepackage[dvips]{graphicx}
\DeclareGraphicsExtensions{.eps}

\newcommand{\be}{\begin{equation}}
\newcommand{\ee}{\end{equation}}
\newcommand{\bea}{\begin{eqnarray}}
\newcommand{\eea}{\end{eqnarray}}
\newcommand{\nn}{\nonumber}

\begin{document}
\bibliographystyle{apsrev}

\title{Nonlinear resonant tunneling of Bose-Einstein condensates in tilted optical lattices}
\author{K. Rapedius}
\affiliation{Department of Physics, Technische Universit\"at Kaiserslautern, 
67653 Kaiserslautern, Germany}
\affiliation{Center for Nonlinear Phenomena and Complex Systems,
Universite Libre de Bruxelles, Code Postal 231, Campus Plaine, 
1050 Brussels, Belgium}
\author{C. Elsen}
\affiliation{Department of Physics, Technische Universit\"at Kaiserslautern, 
67653 Kaiserslautern, Germany}
\author{D. Witthaut}
\affiliation{Network Dynamics Group,
Max-Planck-Institute for Dynamics and Self-Organization,
37073 G\"ottingen, Germany}
\author{S. Wimberger}
\affiliation{Insitut f\"ur theoretische Physik, Universit\"at Heidelberg, 
69120 Heidelberg, Germany}
\author{H. J. Korsch}
\affiliation{Department of Physics, Technische Universit\"at Kaiserslautern, 
67653 Kaiserslautern, Germany}
\date{\today }

\begin{abstract}
We study the tunneling decay of a Bose-Einstein condensate out of tilted 
optical lattices within the mean-field approximation. We introduce a novel
method to calculate also excited resonance eigenstates of the 
Gross-Pitaevskii equation, based on a grid relaxation procedure with 
complex absorbing potentials. This algorithm works efficiently
in a wide range of parameters where established methods fail. It allows
us to study the effects of the nonlinearity in detail in the regime
of resonant tunneling, where the decay rate is enhanced by resonant 
coupling to excited unstable states. 
%Furthermore it allows us to study  for the first time the effect of the 
%nonlinearity on excited states and the emergence of a bistability of the 
%resonance peaks.
\end{abstract}

\maketitle

\section{Introduction}

The dynamics of a quantum particle in a periodic potential subject to 
an external force is one of the central problems in solid state physics. 
In the field free case all eigenstates are delocalized over the lattices, 
leading to transport \cite{Bloc28,Zene32}. The application of a constant 
force leads to a localization of the eigenstates such that transport is 
suppressed contrary to our intuition \cite{Zak68,Wann68,Zak69,Avro77}. 
Instead, the quantum particle performs the celebrated Bloch oscillations, 
and eventually decays by repeated Zener tunneling to higher Bloch bands 
\cite{Daha96,Ande98,Holt00,Gluc00,Gluc02,Jona03,Wimb05,Sias07,Zene08,Gust08,Zene09}. 
The most detailed studies of Bloch oscillations and decay have been 
carried out with ultracold atoms trapped in optical lattices. These systems 
are particularly appealing, because the dynamics of the atoms can be 
recorded in situ and all parameters can be tuned precisely over a wide 
range. The external force can be induced by gravity \cite{Ande98}, 
magnetic gradient fields \cite{Gust08} or by accelerating the lattice 
\cite{Daha96,Jona03,Sias07,Zene08,Zene09}. Decay in strong fields 
manifests itself in the pulsed output of coherent matter waves. 
The dynamics is even more interesting when the atoms undergo 
Bose-Einstein condensation and interactions have to be taken into 
account. For low temperature and high densities, the dynamics of 
the atoms can be described by the celebrated Gross-Pitaevskii 
equation (GPE) with astonishing accuracy \cite{Peth08}. In this 
treatment, interactions are incorporated by a nonlinear mean-field 
potential, which is proportional to the condensate density. The 
nonlinearity of the equation alters the dynamics and in particular 
the decay substantially. Interactions can lead to a damping of 
Bloch oscillations \cite{04bloch_bec}, asymmetric Landau-Zener 
tunneling \cite{Liu02,Jona03,Witt06}, or a bistability of resonance curves 
\cite{Paul05,06nl_transport,Rape08}.

Here we study the resonance eigenstates of the GPE
\bea
  \left( \frac{-\hbar^2}{2m} \frac{d^2}{dx^2} + V(x) + Fx
    + g |\psi(x)|^2 \right) \psi(x) \qquad && \nn \\
      = (\mu -i \Gamma/2) \psi(x)  &&
      \label{eqn-gpe}
\eea
with a periodic potential $V(x+d) = V(x)$ and a static force $F>0$, 
which is known as a Wannier-Stark (WS) potential. The imaginary part 
$\Gamma$ of the eigenenergy gives the decay rate of the condensate. 
In the following we use scaled units with $\hbar = m = 1$ and we 
consider a cosine potential $V(x) = \cos(x)$ unless otherwise stated.
A comprehensive review of the localized eigenstates, the 
WS resonances, can be found in \cite{Gluc02}.

In this article we introduce a new algorithm for the computation
of nonlinear resonance states based on a grid relaxation method
with a complex absorbing potential (CAP). This algorithm converges 
in a wide parameter range and is applicable even to situations 
of many degenerate energy levels, such as the WS system at resonance 
condition (see below). It is thus capable to describe genuine nonlinear 
phenomena such as bistability, which pose a major difficulty to other 
methods as for instance nonlinear complex scaling (CS) 
\cite{Mois05,Schl06,Wimb06,Wimb07,Witt07}.
In addition, it is more efficient and easier to implement and, unlike previous 
methods, is not restricted to ground state calculations but can also compute 
excited states. Note that our approach differs from the CAP 
method used in \cite{Mois04,Schl06} because the latter 
does not use a grid relaxation but relies on a basis set expansion. Though 
such expansions work well for simple single well potentials, they cannot 
easily handle complicated problems like the Wannier-Stark system 
studied in the present paper, which requires the use of as much as 
500 basis states even in the linear (noninteracting) case \cite{Gluc98}.
Our method is applied to study the decay of a Bose-Einstein condensate 
in the strongly nonlinear regime. Nonlinear effects are crucial in the regime 
of resonantly enhanced tunneling (RET). In this case a metastable WS 
resonance becomes energetically degenerate with an excited, less stable 
state, which can increase the decay rate by orders of magnitude.
This phenomenon is most pronounced in deep optical lattices and has 
been studied systematically for the linear case in \cite{Gluc00,Gluc02}. 
The nonlinearity shifts the resonance and eventually bends the 
resonance peak leading to a bistable behavior.

\section{Computational method}

Linear WS resonances can be efficiently calculated 
with the truncated shift operator technique introduced in \cite{Gluc98}. 
In the nonlinear case, the method of CS has been applied 
\cite{Mois05,Wimb06,Wimb07,Witt07}.  
Though satisfactory from a conceptual point of view, this method
has several drawbacks. The implementation is complicated as it 
requires switching between different basis sets as well as different 
time propagation methods. Furthermore, the calculation of excited
states is highly non-trivial, as the method relies on an imaginary
time propagation, and the convergence is quite slow, especially 
for weak fields and close to energetic degeneracies as present 
in the RET condition \cite{Wimb06,Wimb07}.
As an alternative, we propose a method based on complex 
absorbing potentials (CAP) performed on a finite grid 
$[x_-,x_+]$ in real space. We assume that the 
resonance wave function is mainly localized in the interval
$[x_\ell,x_r]$ with $x_- < x_\ell < x_r < x_+$ and
fix the normalization as $\int \nolimits_{x_\ell}^{x_r} |\psi(x)|^2 dx = 1$.
For $x \rightarrow - \infty$, we apply a CAP of the type
\be
  V_{\rm CAP} \propto \left\{ \begin{array}{c l l}
     -i (x/x_-)^{10} & & x < x_- \\
     0 & & x > x_-
     \end{array} \right.  ,
\ee
which only modifies the wave function
in the vicinity of the grid boundary $x_-$ making it square
integrable. For $x \rightarrow + \infty$, the wavefunction rapidly 
converges to zero, so that no further CAP is needed on this
side. The boundary conditions for the wave function read
\be
   \psi(x_-) = 0, \quad \psi(x_+) = 0, \quad \psi'(x_+) = C,
\ee
where the last condition is used to control the normalization.
The algorithm starts from the linear case $g=0$, for which
all WS resonances can be computed efficiently \cite{Gluc98}. 
Nonlinear WS resonances in different bands are calculated by 
choosing a different initial guess.
The nonlinearity is then increased gradually, using the previous
result as initial guess for a standard boundary value problem
(BVP) solver, e.g. the \textsc{Matlab\ }-function {\tt bvp4c}. Applying the BVP
solver changes the normalization of  $\psi$, such that the parameter 
$C$ has to be adjusted according to
\be
   C \rightarrow C \bigg/ 
    \left( \int \nolimits_{x_\ell}^{x_r} |\psi(x)|^2 dx \right)^{1/2}.
\ee
This is repeated until the normalization converges to unity.
Then the nonlinearity is increased by one step.
To demonstrate the validity of this algorithm we compare the
calculated decay rates for a cosine potential for several parameters to complex 
scaling results, which themselves were tested against a direct
time propagation in Ref.~\cite{Wimb06}. The values summarized in 
Tab.~\ref{tab-cs-cap} show an excellent agreement over 
the entire parameter range.
Residual numerical errors are very small; they can mainly be 
attributed to the limited computation time for the CS method
and reflections of the matter wave at the CAP.
For a further discussion of CAPs in the
simulation of few boson systems, see \cite{Lode09} 
and references therein.

\begin{table}[tb]
\centering
\begin{tabular}{p{1.6cm} p{1.6cm} p{2.6cm} p{2.2cm}}
\hline \hline
 $g$ & $F$ &
  $\Gamma_{\rm CS} \quad \quad$ & $\Gamma_{\rm CAP} \quad \quad$ \\
  \hline
  0    & 0.5  & 1.941 $\times \, 10^{-2}$   & 1.941 $\times \, 10^{-2}$ \\
  0.1  & 0.5  & 2.180 $\times \, 10^{-2}$   & 2.180 $\times \, 10^{-2}$ \\
  0    & 0.25 & 7.2 $\times \, 10^{-4}$     & 7.104 $\times \, 10^{-4}$\\
  0.1  & 0.25 & 8.4 $\times \, 10^{-4}$     & 8.346 $\times \, 10^{-4}$\\
  0.2  & 0.25 & 9.7 $\times \, 10^{-4}$     & 9.688 $\times \, 10^{-4}$\\
  0.25 & 0.25 & 1.04 $\times \, 10^{-3}$    & 1.041 $\times \, 10^{-3}$\\
  0.5  & 0.25 & 1.48 $\times \, 10^{-3}$    & 1.476 $\times \, 10^{-3}$\\
  0.2  & 0.15 & 2.9 $\times \, 10^{-5}$     & 2.832 $\times \, 10^{-5}$\\
  0.2  & 0.13125 & 5.7 $\times \, 10^{-5}$  & 5.600 $\times \, 10^{-5}$\\
  \hline \hline
\end{tabular}
\caption{Decay rates $\Gamma$ for the most stable resonance of 
the potential $V(x)=\cos(x)$, taken from Ref. \cite{Wimb06} (CS method) 
and computed by the CAP grid relaxation method.
Particularly for small decay rates the new CAP method 
proves more efficient than the CF technique.
}
\label{tab-cs-cap}
\end{table}

\section{Resonantly enhanced tunneling}

We use the CAP method to investigate how a nonlinear interaction affects the decay of a BEC in a tilted
optical lattice. In the weakly interacting regime, the  
scaling of the decay rate with the field strength is given by the 
celebrated Landau-Zener formula 
$\Gamma(F) \approx F \exp(\pi \Delta E^2/F)$, where $\Delta E$
is the energy gap between the Bloch bands of the periodic
potential \cite{Zene32,Holt00} and the field strength $F$ determines 
the oscillation frequency in the bands \cite{Holt00,Gluc02}. 
Major differences arise in the regime of RET. In this case an 
eigenstate localized mainly in one of the wells of the potential 
becomes energetically degenerate with an exited state in another 
well, which can increase the decay rate by orders of magnitude 
\cite{Gluc02}. In the following, we focus on the experimentally 
studied regime \cite{Sias07,Zene08}, where already a modest nonlinearity strongly affects the
decay of the condensate \cite{Wimb05,Sias07,Zene08}.

\begin{figure}[tb]
\centering
\includegraphics[width=8.5cm,  angle=0]{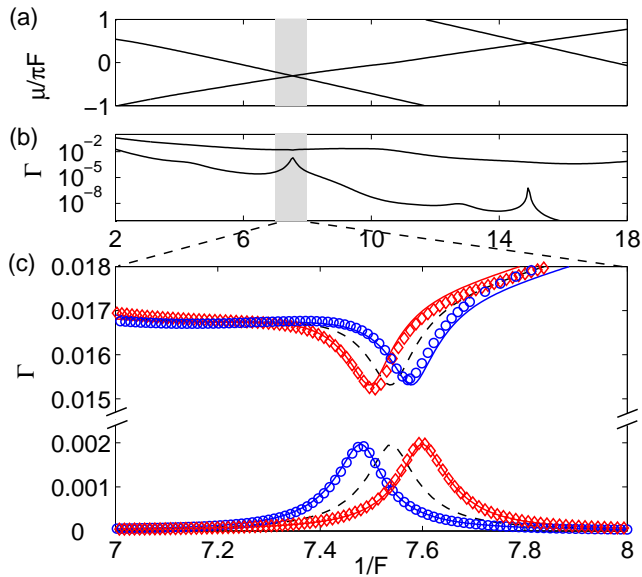}
\caption{\label{fig-retpeaks}
(Color online) Resonantly enhanced tunneling (RET) of (non)-linear
WS resonances.
(a) Energies and (b) decay rates of the two most stable WS resonances 
in a cosine-potential as a function of the inverse field strength $1/F$. 
(c) Shift of the RET peaks due to the nonlinear interaction of a BEC
for $g=+0.02$ ($\circ$), $g=-0.02$ ($\diamond$) and $g=0$ (- -).
Numerical results (symbols) are compared to a perturbative 
calculation (solid lines) according to 
Eqs.~(\ref{eqn-cnl-deltae}) and (\ref{egn-gamma-pert}).
 }
\end{figure}

RET is illustrated in Fig.~\ref{fig-retpeaks} 
(a,b) for the linear case $g=0$, showing the decay rate $\Gamma$ and 
the chemical potential $\mu$ of the two most stable resonances as a 
function of $F$. RET is observed at $1/F \approx 7.5$, 
where the two energy levels $\mu(F)$ cross. The resonant coupling to
the excited states leads to a pronounced RET peak of the decay rate
for the most stable resonance. Coincidentally, a pronounced dip
is observed for the first excited resonance, which is stabilized by the
coupling to the most stable resonance \cite{Gluc02}.
The influence of a small nonlinearity is illustrated in 
Fig.~\ref{fig-retpeaks} (c). Three main effects are observed: a shift 
of the resonance peaks, an increase (decrease) of the peak decay 
rate in the ground state for $g>0$ ($g<0$) and a deformation of the 
peak shape. 

\begin{figure}[tb]
\centering
\includegraphics[width=8.5cm,  angle=0]{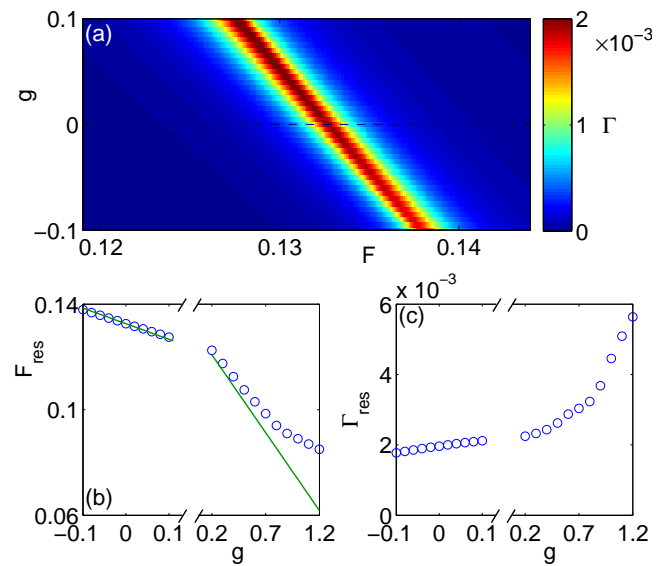}
\caption{\label{fig-peakvsg}(Color online)
(a) Colormap plot of the decay rates of the most stable WS 
resonances in a cosine-potential vs. the field 
strength $F$ and the interaction strength $g$ in the
vicinity of the first order RET peak. (b) Position and (c) height 
of the RET peak vs. the interaction strength $g$.
}
\end{figure}

The shift and the deformation can be qualitatively 
understood by a perturbative approach \cite{Wimb06}. 
To first order, this predicts a shift of the real part of the eigenenergy, 
\be
 \Delta \mu(g) \approx g \int_{x_\ell}^{x_r}  |\psi_{g}|^2|\psi_{g=0}|^2 dx
   \approx g \int_{x_\ell}^{x_r}  |\psi_{g=0}|^4 dx,
   \label{eqn-cnl-deltae}
\ee
which corresponds to a shift of the field strength
according to
\be
   \Delta F(g) \approx \pm \Delta \mu(g) /(2\pi).
   \label{eqn-pert-f}
\ee
Here, the minus sign holds for the ground and 
the plus sign for the excited band.
The nonlinear decay rate is then approximately given by 
\be 
  \Gamma_g(F) = \Gamma_0(F + \Delta F(g)).
  \label{egn-gamma-pert}
\ee

The shift is further investigated in Fig.~\ref{fig-peakvsg} (b), where the 
decay rate as well as the peak position is plotted vs.
the interaction strength over a wide parameter range.
The perturbative calculation (\ref{eqn-cnl-deltae}) predicts 
that the peak position $F_{\rm res}$ is shifted with a slope 
$d F_{\rm res}/dg = 0.059$ for small values of $g$, which is 
plotted as a green line in Fig.~\ref{fig-peakvsg} (b). 
This deviates from the numerically exact results
already for small values of $g$, 
for which a linear fit yields a smaller slope of
$d F_{\rm res}/dg = 0.051$.
In agreement with \cite{Wimb06} we thus find that
first order perturbation theory is not sufficient to describe
the shift of the RET peaks quantitatively.
Noticeably, the RET peak and the dip of the decay rate for the 
first excited resonance always shift into opposite directions, as
shown in Fig.~\ref{fig-retpeaks} (c).

The change in the maximum decay rate is not predicted by 
perturbation theory, but easily explained phenomenologically.
It is a direct consequence of the interaction as repulsion between 
the particles in general leads to a destabilization whereas attraction 
leads to a stabilization of both resonances and bound states 
(see \cite{Jona03} and references therein). 
This is further illustrated in Fig.~\ref{fig-peakvsg} (c), where the 
peak decay rate of the most stable resonance is plotted as a 
function of  $g$ over a wide parameter range.
Similar effects have been investigated for several other model 
potentials \cite{Mois05,Wimb07,06nl_transport}.

\begin{figure}[tb]
\centering
\includegraphics[width=7.5cm,  angle=0]{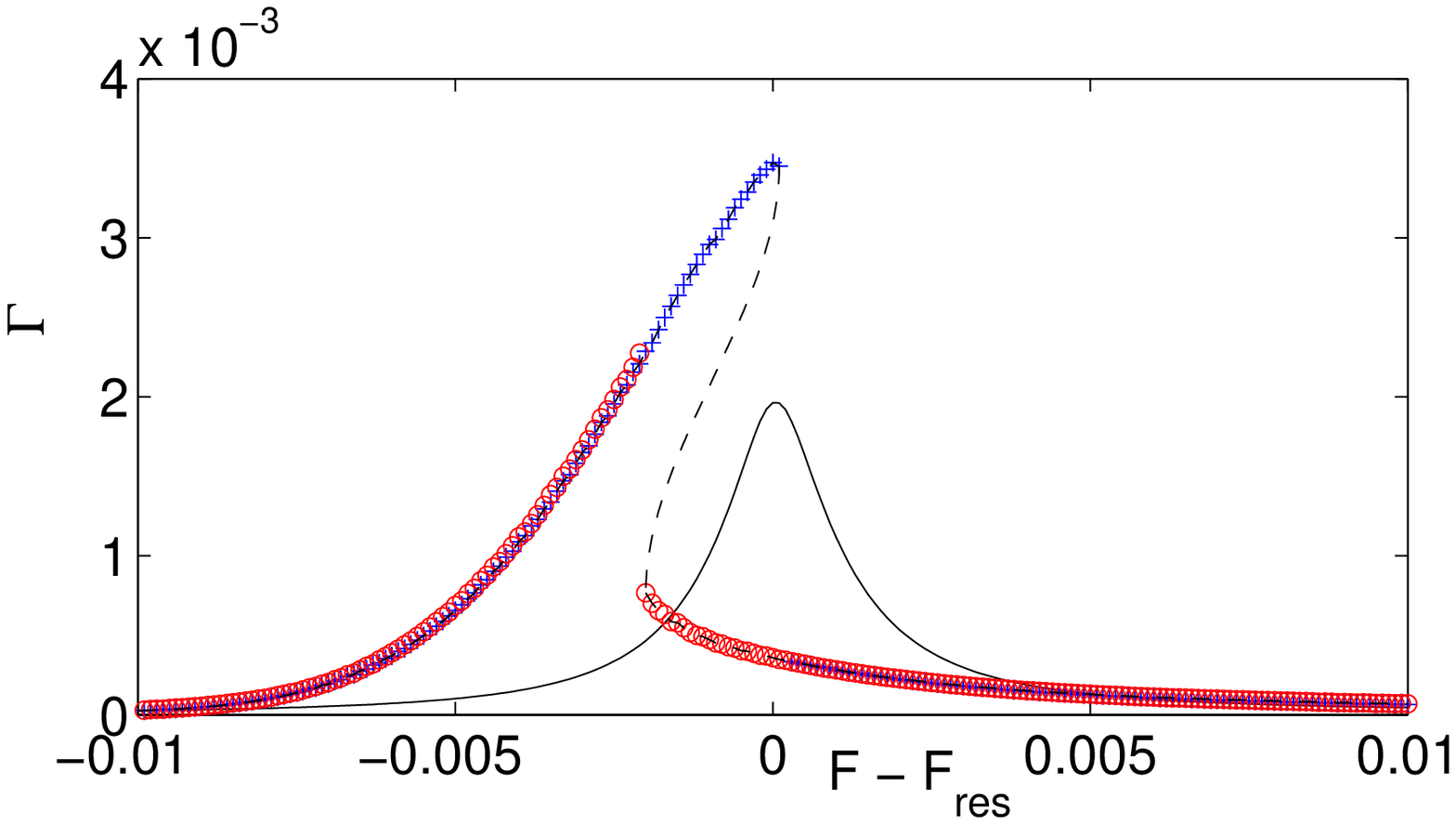}
\includegraphics[width=7cm,  angle=0]{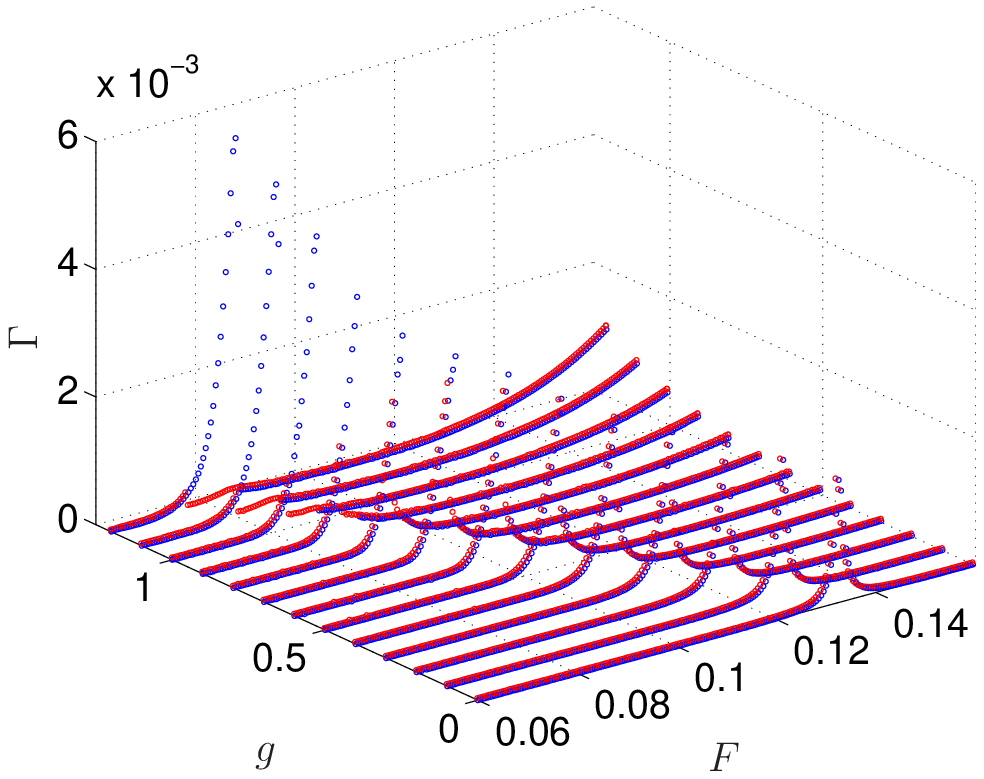}
\caption{\label{fig-bistability} (Color online)
Upper panel: Bistability of the RET peak for strong repulsive interactions
($g = 0.8$). The decay rate was calculated for a forward 
sweep (blue asterisk) and a backward sweep (red circles). A 
spline interpolation (dashed line) is included to guide
the eye. The solid line shows the linear ($g=0$) peak shape for 
comparison.
Lower panel: Emergence of the bistable behavior: Decay rate as 
a function of the field strength $F$ for different values of the nonlinearity
$g$.
 }
\end{figure}

Figure \ref{fig-retpeaks} (c) furthermore shows that the RET peaks
become asymmetric for $g \neq 0$. For a repulsive (attractive) 
nonlinearity, the peak bends to higher (lower) values of $F$. 
If the nonlinearity is increased above a critical value $g_{\rm cr}$, 
the peaks bend over and a bistable behavior emerges as shown 
in Fig.~\ref{fig-bistability}. For the given parameters 
of the optical lattice, the critical nonlinearity has been found at
$0.5 < g_{\rm cr} < 0.6$. 
The detailed shape of a bistable RET peak is plotted in the upper
panel of Fig.~\ref{fig-bistability}, which also indicates how  
WS states are calculated numerically in the bistable regime:
We have started with a small value of $F$ which was then gradually 
increased, using every result as initial guess for the next calculation
(cf. also \cite{Wimb06,Wimb07}). After reaching a final, large 
value of the field strength, the procedure was reversed and $F$ 
was decreased back to the initial value. Within the regime of
bistability forward and backward sweep yield the upper and lower
branches of the peak, respectively. The intermediate branch is 
generally difficult to compute as it is dynamically unstable.

The bending of the RET peak and the emergence of a bistability
can be understood qualitatively by the perturbative approach
introduced above. 
A common WS state in a deep optical lattice is strongly localized 
in a single potential well so that its chemical potential is strongly
changed according to Eq. (\ref{egn-gamma-pert}).
In comparison, the state corresponding to the maximum of the RET 
peak is delocalized because of the energetical degeneracy 
with an excited state in another well. Therefore its chemical 
potential is affected rather weakly and according to 
Eq. (\ref{eqn-pert-f}) also the change of the peak position 
$\Delta F$ is small (cf. also \cite{Sias07,Zene08}).
With increasing nonlinearity, the edges of a RET peak shift to 
smaller values of the field strength, while the maximum falls 
behind. The whole peak bends to the right and finally becomes 
bistable.

Bifurcations of nonlinear stationary states have been previously 
observed in several different contexts: The simplest example 
is the nonlinear two-mode system. As observed for the WS state, 
new stationary states emerge first in the vicinity of the resonance, 
i.e. the when the chemical potential of the two coupled states 
coincides (see \cite{Liu02,Witt07} and references therein).
The emergence of bistability has also been analyzed for
the transmission coefficient in the context of nonlinear RET 
through one-dimensional potential barriers 
\cite{06nl_transport,Paul05,Rape08}.
However, in this case states corresponding to the transmission
maximum are localized strongest. Thus the resonance peaks bend 
into the same direction as they are shifted, which is opposite
from the behavior of the WS RET peaks shown in 
Fig.~\ref{fig-bistability}.

\begin{figure}[tb]
\centering
\includegraphics[width=8.5cm,  angle=0]{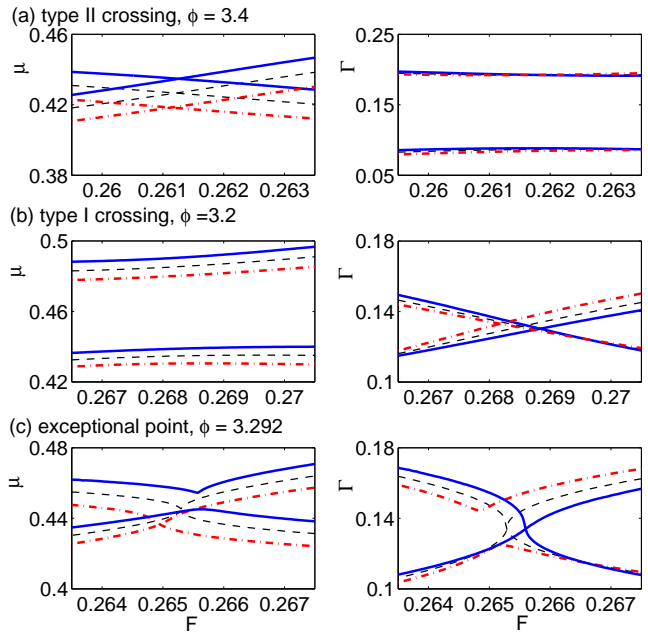}
\caption{\label{fig-exceptional}
(Color online)
Chemical potential $\mu$ and decay rates $\Gamma$ for (non)linear
WS-resonances in a bichromatic optical lattice for $\delta = 1$ and
(a) $\phi = 3.4$, (b) $\phi = 3.2$, (c) $\phi = 3.292$ and
$g=0$ (dashed black line), $g=+0.02$ (thick blue line) and
$g=-0.02$ (dash-dotted red line).
}
\end{figure}

\section{Beyond the RET-regime}

A new regime of RET can be explored in
bichromatic optical lattices, 
\be
   V(x) = \frac{1}{2} \cos(x) + \frac{\delta}{2} \cos(2x + \phi).
\ee
These potentials can be realized experimentally by superimposing
two incoherent optical lattices \cite{Gorl01,Foll07}, or by combining optical 
potentials based on virtual two-photon and four-photon processes
\cite{Ritt06,Salg07}. It can be used to study Landau-Zener tunneling
between different bands and the interplay of tunneling
and Bloch oscillations \cite{Brei07}.

Depending on the relative phase and the relative strength of the
two lattices, WS resonances in tilted bichromatic optical lattices
show two remarkably different types of level crossing scenarios 
\cite{Keck03} -- either the real parts $\mu$ (type-I) or the imaginary
parts $\Gamma$ (type-II) of the eigenenergies cross. A full degeneracy 
of both $\mu$ and $\Gamma$ occurs only for isolated points 
in parameter space. Examples are shown in Fig.~\ref{fig-exceptional} 
for $\delta = 1$ and different relative phases $\phi$ of the two 
lattices. A familiar type-II crossing is observed for $\phi = 3.4$. The real 
parts of the eigenenergies ($\mu$) cross, while the imaginary
parts $\Gamma$ anti-cross, leading to the familiar RET-peaks
of the decay rates. Changing the phase slightly to $\phi = 3.2$, one finds a 
type-I crossing. The decay rates of the two most stable 
resonances now cross, while the {\it real} parts anti-cross.
An exceptional point \cite{Keck03}, where both real and imaginary part
are fully degenerate, is found for $\phi = 3.292$, as shown
in Fig.~\ref{fig-exceptional} (c). However, the degeneracy
is lifted as soon as the atoms start to interact. 
A weak repulsive nonlinearity $g=+0.02$, turns the exceptional
crossing into an ordinary type-I crossing, while an attractive
nonlinearity $g=-0.02$ favors a type-II crossing.
This change of peak shape can have dramatic effects on the
dynamics of a Bose-Einstein condensate, in particular when
experimental parameters are adiabatically varied 
(see, e.g.~\cite{Keck03}).

\section{Conclusions}

Bose-Einstein condensates in tilted optical lattices are ideal
to study the decay of interacting open quantum systems. 
Experimentally the parameters can be tuned over a wide
range and the dynamics can be recorded in situ.
Here we presented an {\em efficient} method
to calculate the decay rate in the mean-field regime also
in the presence of degeneracies which also, unlike previous 
methods, is not restricted to ground state calculations. 
The effects of the nonlinearity are strongest in the regime
of resonant tunneling, where the decay rate can be enhanced
by orders of magnitude by resonant coupling to unstable
excited states. The interactions shift and bend the resonance 
peaks and eventually lead to a bistable peak shape. 
Even more interesting effects can be studied in tilted
bichromatic lattices, where different types of level
crossing scenarios emerge when the lattice parameters 
are tuned. These effects will be studied in detail in 
a future publication.

\acknowledgements

Financial support by the Deutsche Forschungsgemeinschaft (DFG) 
via the Graduiertenkolleg 792, the Heidelberg Graduate School 
of Fundamental Physics (grant number GSC 129/1) and the
Forschergruppe 760 (grant number WI 3426/3-1) is gratefully 
acknowledged.
K. R. receives financial support 
as 'boursier of the Universit\'e Libre de Bruxelles'.

% --- Literatur -------------------------------------------------------------------

\end{document}